# Generative AI in Cybersecurity

Shivani Metta[1], Isaac Chang[2], Jack Parker[3], Michael P. Roman[1], & Arturo F. Ehuan[1*]

[1]Cybersecurity Master of Engineering Program, Duke University, Durham, NC 27708, USA

[2]Lynbrook High School, San Jose, CA 95129, USA

[3]Pratt School of Engineering, Duke University, Durham, NC 27708, USA

*Lead Author. Corresponding Author. Cybersecurity Master of Engineering Program, Duke University, Durham, NC 27708, USA. E-mail: ehuan.arturo@duke.edu

# Abstract

The dawn of Generative Artificial Intelligence (GAI), characterized by advanced models such as Generative Pre-trained Transformers (GPT) and other Large Language Models (LLMs), has been pivotal in reshaping the field of data analysis, pattern recognition, and decision-making processes. This surge in GAI technology has ushered in not only innovative opportunities for data processing and automation but has also introduced significant cybersecurity challenges.

As GAI rapidly progresses, it outstrips the current pace of cybersecurity protocols and regulatory frameworks, leading to a paradox wherein the same innovations meant to safeguard digital infrastructures also enhance the arsenal available to cyber criminals. These adversaries, adept at swiftly integrating and exploiting emerging technologies, may utilize GAI to develop malware that is both more covert and adaptable, thus complicating traditional cybersecurity efforts.

The acceleration of GAI presents an ambiguous frontier for cybersecurity experts, offering potent tools for threat detection and response, while concurrently providing cyber attackers with the means to engineer more intricate and potent malware. Through the joint efforts of Duke Pratt School of Engineering, Coalfire, and Safebreach, this research undertakes a meticulous analysis of how malicious agents are exploiting GAI to augment their attack strategies, emphasizing a critical issue for the integrity of future cybersecurity initiatives. The study highlights the critical need for organizations to proactively identify and develop more complex defensive strategies to counter the sophisticated employment of GAI in malware creation.

# Introduction/Foreword

## 1.1. Introduction

Generative Artificial Intelligence (GenAI) has opened a new frontier for defensive and offensive cyber operations in the ever-evolving cybersecurity landscape. While generative AI holds immense potential for innovation and efficiency, its capabilities also present unprecedented challenges, particularly in the development and enhancement of malware. This paper delves into the intricate dynamics of how cyber adversaries are leveraging generative AI to craft sophisticated, adaptive, and elusive malware, posing significant threats to cybersecurity infrastructures globally.

The rapid advancements in AI technologies, especially those related to machine learning (ML) and deep learning (DL), have significantly broadened the scope of what is achievable regarding automated system attacks and defenses. ML refers to algorithms that can learn from data without explicit programming, while DL builds upon ML by utilizing complex artificial neural networks to learn from intricate patterns within data. This progress, however, presents a double-edged sword. On one hand, AI has enhanced cybersecurity measures, enabling the development of robust, predictive security systems. On the other, it has equally empowered cyber adversaries, who utilize these technologies to develop malware that can outsmart traditional security measures , adapt to new environments, and evade detection with alarming efficiency.

This paper explores the various facets of this emerging threat. It will provide an in-depth analysis of the evolution and trends of generative AI in the context of cybersecurity, focusing on its use in the development and enhancement of malware. Questions arising from this include:

- How is Generative AI reshaping the landscape of cybersecurity, and what roles does it play in both defense and offense?
- Is the cybersecurity industry witnessing a pivotal shift towards employing Generative AI for malicious purposes, and what could this mean for the future of digital security?

Other sections will examine the development and effectiveness of generative AI for malicious purposes, as well as the potential risks and impacts of these technologies, and the challenges they pose to cybersecurity professionals.

- How does Generative AI compare to traditional human methodologies in terms of accelerating malware creation, and what are the implications of this technological evolution?
- Which industry segments are most at risk from the threats posed by Generative AI-driven malware, and why are they particularly vulnerable?
- Who are the cyber adversaries most likely to adopt Generative AI for their activities, and what drives their choice of this advanced technology?
- What can a detailed case study of an AI-augmented malware assault reveal about the evolving capabilities of AI to create or refine hostile software?

Finally, the paper will address the ethical implications of using GenAI and propose strategies to mitigate the risks associated with AI-enhanced malware. Through a comprehensive review of literature and current events, the paper aims to offer insights into the future trajectory of cyber threats and the evolving role of AI in cybersecurity.

- What are the significant dangers and consequences associated with the use of AI in cyber conflict, and how do these challenges redefine cybersecurity paradigms?
- What moral dilemmas arise from using AI in cyber warfare, and what array of strategies exist to effectively counter these AI-related threats?
- What does the future hold for combating AI-enhanced cyber threats, and what actionable guidance can be offered to those tasked with safeguarding our digital frontiers?

The paper seeks to contribute to the broader understanding of generative AI's dual use in cyber operations and to foster a more informed approach to combating AI-enhanced cyber threats.

# 2. Navigating the Evolution and Unfolding Trends

## 2.1. Artificial Intelligence evolution

Since the mid-20th century, there have been many phases of development in artificial intelligence, such as Rule-based systems, Machine Learning, Deep Learning and Generative AI.
- *Rule-based systems:* These early AI systems relied on pre-programmed rules and logic to solve specific problems. While limited in their flexibility, they laid the foundation for future developments. [1]
- *Machine Learning (ML)*: ML systems learn from data without explicit programming, achieving superior performance in tasks like pattern recognition and prediction. This marked a major step towards more versatile AI capabilities.[1]
- *Deep Learning (DL)*: A subfield of ML, DL employs artificial neural networks inspired by the human brain, enabling the processing of complex data like images, text, and audio. DL has driven breakthroughs in areas like natural language processing and computer vision.[1]
- *Generative AI (GenAI)*: This latest iteration of AI utilizes ML algorithms to generate entirely new content, such as text, audio, or video, based on user input. GenAI has rapidly gained traction due to its ability to mimic various creative processes [2].

Generative AI (GenAI) leverages machine learning algorithms to create entirely new content, such as text, audio, or video, based on user input. One of the most prominent approaches within GenAI are Large Language Models (LLMs). These powerful models are trained on massive amounts of text data, allowing them to analyze and understand language patterns effectively[3].
LLMs, like the recently released GPT-4 (Generative Pre-trained Transformer) by OpenAI, are at the forefront of GenAI advancements. These models can generate human-quality text, translate languages, write different kinds of creative content, and even answer your questions in an informative way [4].
Similar to other technologies, GenAI presents a double-edged sword. While it holds immense potential for positive applications in various fields, it also raises concerns about potential misuse. It's crucial to acknowledge both sides of this powerful technology as we move forward.

On the offensive side of the house, exploiting any systems using malware is the main concern. Currently, hackers need to be proficient in understanding the underlying technology, business use cases, and detective & preventive controls to hack into the systems. With the help of GenAI, script kiddies will also be able to match the proficiency of hackers to create the malware. Recently, the market witnessed a surge in AI models specifically optimized and trained for malware development, marking a significant shift in the methodologies employed by cybercriminals. While malware creation traditionally required substantial technical expertise, advances with large language models (LLMs) have drastically lowered this barrier. Individuals with minimal technical skills can generate sophisticated malware by providing skillful prompts to these publicly available AI systems [5]. This development broadens the potential pool of threat actors. It poses a significant challenge to cybersecurity defenses, necessitating reevaluating security strategies to mitigate these types of threats.

## 2.2. Impacts of GenAI on cybersecurity

Generative Artificial Intelligence (GenAI) has taken a significant place in the cybersecurity sector due to its proactive capabilities. While many cybersecurity vendors use "AI" as a marketing buzzword, the full potential of GenAI, particularly large language models (LLMs), is beginning to revolutionize the industry.Aspiringly these tools are being trained on extensive historical cybersecurity data, to be capable of identifying emerging patterns and trends to predict future threats, thereby enabling cybersecurity professionals to adopt a more anticipatory stance against potential breaches and to help enhance the efficiency and effectiveness of their existing security measures [6]. This would enable enterprises to secure their systems more effectively by generating complex and unique passwords or encryption keys, making them very difficult for cyber attackers to guess or crack [7]. This aspect of GenAI can vastly improve an organization's security posture by enhancing the robustness of its first line of defense against unauthorized access.

In terms of strategic investment, organizations recognize the importance of GenAI in cybersecurity, with many planning to increase their AI investment to harness the advancements with this technology [8]. This suggests a growing reliance on GenAI technologies to bolster cyber defenses and potentially transform how cybersecurity is approached at an organizational level.

Generative AI is specifically praised for strengthening cybersecurity through improved threat identification. While it may not lead to a completely automated system, it aids in the rapid and accurate detection of potential cyber threats [9]. LLM-based GenAI shows significant promise in strengthening cybersecurity by improving threat identification. These advanced models can analyze vast amounts of data, including historical security breaches and ongoing threat intelligence, to detect potential cyber threats with greater accuracy and speed than manual methods. This empowers proactive security measures by identifying vulnerabilities and predicting future attacks before they occur.

However, it's important to acknowledge that complete automation remains an aspiration in the near future. While LLM-based GenAI can significantly augment and enhance the work of cybersecurity professionals, it cannot completely replace their expertise and judgment.
Here's how LLM-based GenAI supports cybersecurity teams:
- Rapid and accurate threat detection: LLMs can analyze massive datasets to identify anomalies and suspicious patterns, significantly reducing the time and resources needed to detect potential threats[10].
- Predictive capabilities: By learning from historical data and ongoing threat intelligence, LLMs can predict future attack vectors and vulnerabilities, enabling proactive security measures[10].
- Improved efficiency and productivity: GenAI can automate tedious tasks like threat analysis and incident response, freeing up valuable time for human experts to focus on strategic decision-making and complex investigations [11].

Looking forward, the responsible development and implementation of LLM-based GenAI will be crucial in bolstering cybersecurity defenses . However, cautious optimism is necessary as we navigate the gap between aspirational marketing claims and the current realities of AI technology in cybersecurity.

Moreover, these advanced AI models can learn from previous security breaches to predict and detect future violations much more quickly than manual monitoring could achieve. This predictive capability is invaluable in a landscape where threats evolve rapidly and often outpace traditional reactive security measures.

### 2.2.1. Offensive Usecases and the Attribution Dilemma

Beyond the defensive potential of GenAI lies a dark side: its formidable offensive capabilities. Malicious actors have always been a step ahead of those who need to defend against them. In the wrong hands, this technology can be weaponized to launch devastating cyberattacks, posing a grave threat to individuals, organizations, and national security.By being aware of the potential dangers and implementing robust safeguards, we can mitigate the risks associated with GenAI's offensive capabilities and ensure its responsible use in the realm of cybersecurity.

One of the most alarming applications of GenAI in cyberwarfare is the creation of hyper-personalized and realistic phishing campaigns. Leveraging machine language and natural language processing (NLP), malicious actors can craft personalized emails, text messages, and social media posts that appear to originate from trusted sources, such as banks, online platforms, or even close friends and colleagues. A presentation at Black Hat 2023 showcased this vulnerability, demonstrating how emails crafted with the help of large language models (LLMs) achieved a concerning 80% click-through rate.[12], This highlights the serious threat posed by such personalized and AI-powered phishing campaigns, as users are more likely to be tricked by messages that appear tailored to them and their specific circumstances.

The development of more potent malware and cyberattacks can also be facilitated by GenAI. By automating the process of code generation and mutation, attackers can create polymorphic malware that readily evades detection by traditional antivirus software. This automation enables rapid creation and deployment of new threats, making it challenging for cybersecurity professionals to keep pace and develop effective countermeasures.

Distributed denial-of-service (DDoS) attacks, which can cripple online services and critical infrastructure, are another potential application of GenAI in the offensive realm. This can be achieved by[13]
- Threat Prediction: LLMs can analyze large datasets of historical attack patterns and identify anomalies, potentially predicting future DDoS attempts before they occur. This allows for proactive measures to be taken, such as scaling infrastructure or implementing additional security measures.
- Automated Response: AI can be used to automate detection and response processes during a DDoS attack. This can include filtering malicious traffic, rerouting requests, and mitigating the impact on legitimate users.

The widespread adoption of GenAI technologies poses a significant challenge in terms of attribution. As these tools become more accessible and sophisticated, identifying the perpetrators of cyberattacks will become increasingly difficult. This poses a major obstacle for law enforcement and cybersecurity professionals, hindering their ability to bring cybercriminals to justice and prevent future attacks.

By recognizing the potential of GenAI as a double-edged sword and implementing proactive strategies, we can harness its defensive capabilities while mitigating the risks associated with its offensive applications. This requires a collaborative effort from researchers, policymakers, and cybersecurity professionals to ensure that GenAI becomes a force for good in the digital world[14] [15].

## 2.3. The growing landscape of GenAI for nefarious use

There is a growing concern about the pivot toward employing Generative AI for malicious purposes. While generative AI has the potential to significantly enhance cybersecurity, particularly in threat identification, it is also being explored by bad actors to aid in cyberattacks. Innovations such as self-evolving malware, which can adapt and change to avoid detection, are particularly concerning [9] [16].

The potential for nefarious use of generative AI in cyberattacks cannot be overstated. As a powerful tool, generative AI can be harnessed for creating sophisticated phishing campaigns, generating content that is free of typos and grammatical errors, thereby increasing the likelihood of deceiving the recipients. This trend toward more manipulative and seemingly legitimate phishing content is a direct result of the misuse of generative AI [16] [17].

Moreover, cyber threat actors may exploit generative AI to create malware that leverages zero-day vulnerabilities, generate malicious websites that appear legitimate, personalize phishing emails, and produce deep fake data to deceive users and systems. The capacity of generative AI to inundate security systems with sophisticated threats is unprecedented, raising the stakes for cybersecurity defenses. Deepfakes are hyper-realistic media manipulations, typically videos, that use AI to replace or alter a person's likeness or voice in existing footage. This technology can be used to create highly believable videos of individuals saying or doing things they never actually did, raising concerns about their potential to spread misinformation and damage reputations. [7][18] Synthetic media encompasses a broader range of AI-generated content, including not just deepfakes but also fabricated audio recordings and images. This technology further blurs the lines between reality and fiction, making it increasingly difficult to discern genuine content from manipulated materials. [8][18]
These fabricated materials can be readily exploited to spread disinformation, manipulate public opinion for malicious purposes, sow discord within communities, and damage the reputations of individuals and organizations. The hyper-realistic nature of these deepfakes and synthetic media makes attribution (identifying the source of the false information) an increasingly challenging task, further adding to the potential harm they can cause.

Generative AI's role in enhancing phishing campaigns has been noted as a particular area of concern. Cybercriminals are utilizing AI to craft more manipulative and successful phishing campaigns, which are often the first step in a multi-stage attack to steal credentials and infiltrate systems. These AI-enhanced phishing attacks demonstrate how generative AI can be weaponized to create high-impact cybersecurity threats [19].

This emerging trend highlights a dual-use dilemma where the same technologies developed to protect and secure can also be repurposed to attack and exploit. The cybersecurity industry must therefore remain vigilant and develop robust countermeasures to mitigate the risks associated with the malicious use of generative AI.

The ability of LLMs to understand and manipulate code makes them potent tools for both attackers and defenders. They can be used to automate complex tasks, analyze vast amounts of data to identify vulnerabilities, and even develop defensive countermeasures against other AI-powered attacks. However, the same capabilities can be used by malicious actors to automate malware generation, create more sophisticated phishing campaigns, and even evade security detection systems. The emergence of LLM models like WormGPT signifies a concerning trend in the cybersecurity landscape, where advanced AI tools are being repurposed for malicious intent. WormGPT, as highlighted in several reports, is a prime example of this. Developed as a private chatbot service, WormGPT is advertised as a tool that uses AI to write malicious software, circumventing the typical restrictions placed on such activities [20].

Based on the GPTJ large language model from 2021, which was developed by EleutherAI, WormGPT is specifically designed for malicious activities. It possesses features that are particularly useful for cybercriminals, such as unlimited character support, the ability to retain chat memory, and efficient code formatting, but the biggest enabling feature would be the removal of security guardrails that are present in ChatGPT. These features make WormGPT a potent tool for developing sophisticated phishing attacks and business email compromise schemes [21] [16].

The existence and accessibility of tools like WormGPT underline the need for the cybersecurity industry to remain vigilant and proactive. As these models become more sophisticated and accessible, the potential for their misuse grows, necessitating enhanced defensive measures and ethical guidelines to prevent the exploitation of AI for harmful purposes.

A case study by CyberArk shows that "GenAI can enable the development of polymorphic malware, posing significant challenges to cybersecurity." Polymorphic malware can alter its code as it spreads, evading signature-based detection methods. By leveraging GenAI, these malware variants can now generate numerous, unique iterations of their code, thereby complicating detection and neutralization efforts [22].

The research from CyberArk demonstrates how dialogue systems like ChatGPT can be manipulated into generating polymorphic malware, showcasing the potential for AI-driven tools to be subverted for creating advanced cyber threats. Similarly, Tripwire's analysis underscores "the sophistication of polymorphic and metamorphic malware, which can not only alter their appearance but also their underlying code—making them particularly elusive to traditional antivirus solutions" [23] [24].

This development signifies a paradigm shift in malware creation and detection, necessitating novel approaches in cybersecurity strategies, including the deployment of advanced heuristic and behavior-based detection mechanisms capable of identifying and mitigating such chameleon-like threats. As GenAI continues to evolve, the arms race between cyber attackers and defenders grows increasingly complex, with the former gaining access to tools that can significantly automate and enhance their malicious capabilities.

## 2.4. Understanding the potential of GenAI and identifying high-risk sectors in cybersecurity

### 2.4.1. Comparison of Traditional to GenAI powered malware development

GenAI has introduced a new dimension to the ongoing battle against cyber threats. While traditionally, malware creation has been a time-consuming and skill-dependent process, GenAI offers a concerning potential for accelerated development, increased variation, and lowered barriers to entry for malicious actors.

Comparing GenAI and traditional methodologies, exploring their strengths and weaknesses in the context of malware development, and delve into the broader implications of this technological evolution.

*Traditional Malware Creation:*
- Time-consuming and Labor-intensive: Crafting effective malware traditionally requires significant time and effort. Developers need to possess strong programming skills and a deep understanding of security vulnerabilities to write, test, and refine their code.
- Limited Variability: Manually modifying code to create diverse malware variants is a complex and time-consuming process, hindering the ability to effectively evade detection methods that rely on signature-based identification.
- Skill-dependent Barrier: The technical expertise needed to create sophisticated malware restricts the pool of potential attackers, limiting the overall threat landscape.[25][26]

*GenAI in Malware Development:*
- Accelerated Development: Trained on vast datasets of existing malware, GenAI models can generate new variants significantly faster than humans. This automation streamlines the development process, potentially leading to a surge in the volume of malware created.
- Enhanced Variation: GenAI can produce highly diverse and novel malware by generating code with specific functionalities and employing obfuscation techniques, making it difficult for traditional detection methods to identify and block these threats.
- Lowered Skill Barrier: User-friendly interfaces and pre-trained GenAI models could potentially lower the technical barrier to entry, allowing individuals with less expertise to engage in malicious activities.[25][26]

*Implications of GenAI-driven Malware:*

- Increased Attack Volume and Complexity: The ability to generate large volumes of diverse malware can significantly increase the number of attacks and their complexity, making them more challenging to detect and prevent.
- Faster Threat Evolution: Rapid development of new malware variants through AI can outpace the development of traditional security solutions, creating a constant struggle for defenders to keep pace with the evolving threat landscape.
- Broader Range of Actors: Lowering the skill barrier for creating malware could lead to a broader range of individuals engaging in cybercrime, potentially leading to more widespread and sophisticated attacks targeting various sectors.[24]

By understanding the strengths and weaknesses of both traditional and GenAI-based approaches to malware creation, we can develop effective countermeasures and leverage AI-powered solutions to enhance defense mechanisms.

## 2.4.2. Identifying potential high-Risk Sectors for GenAI powered malware attack

In 2023, according to the Identity Theft Resource Center's 2023 Data Breach Report malware attack was the third most common cyberattack vector in the United States[26]. The potential dangers of GenAI-driven malware extend to various industries, but some sectors face greater vulnerability due to specific characteristics that make them attractive targets. Here are some of the most at-risk segments:

1. **Financial Services:**

The report states that the financial services industry accounted for 24% of all data breaches in the United States, exposing millions of sensitive records [26].
Financial institutions store sensitive data like user credentials, financial information, and personally identifiable information (PII). This data is highly valuable to attackers for various malicious activities like identity theft, fraud, and financial extortion. They often rely on complex IT infrastructure, including legacy systems, which may have vulnerabilities that GenAI-driven malware can exploit to gain access. The strict regulatory requirements can pressure financial institutions to implement security measures quickly, potentially leading to rushed adoption of new technologies without thorough security testing, making them susceptible to novel threats like GenAI malware.

2. **Healthcare:**

Healthcare organizations rely on critical infrastructure for patient care, making them vulnerable to disruptions caused by malware attacks. Healthcare data is highly sensitive, including patient medical records, which can be used for identity theft, targeted medical scams, or even blackmail. Compared to other sectors, some healthcare organizations might have limited resources to invest in robust cybersecurity measures, making them more susceptible to attacks.

3. **Critical Infrastructure:**

Attacks on critical infrastructure, such as power grids, transportation systems, and communication networks, can have widespread societal and economic disruptions. Critical infrastructure often involves interconnected and complex systems, making them challenging to secure and potentially offering multiple entry points for GenAI-driven malware. Many critical infrastructure systems have limited tolerance for downtime, making them more susceptible to pressure tactics employed by attackers after a successful breach.

   4. **Government Agencies:**

Government agencies store and process vast amounts of sensitive data, including national security secrets, citizen information, and classified intelligence. This data is highly valuable for espionage and foreign interference attempts. Government agencies often have extensive network connections with other organizations and critical infrastructure, potentially expanding the attack surface and amplifying the impact of a successful breach. Government agencies may have complex IT environments with older and potentially unpatched systems, creating vulnerabilities that GenAI-driven malware can exploit.

It's crucial to remember that continuous vigilance and proactive measures are essential to mitigate the risks posed by GenAI-driven malware across all industries.

# 3. Development and effectiveness of GenAI

There are several methods where GenAI may be used to assist cyber adversaries, including by lowering the skill level for cyber criminal activities, and assisting in the creation of malware, and facilitating other malicious activities. The GenAI of choice is ChatGPT 3.5, but other models will also be explored.

## 3.1. Lowering the skill floor for cyber criminals

Developing malware and effective social engineering attacks takes both skill and time - time to learn and time to execute. Online safety awareness and security continues to improve, so cyber criminals must invest time to learn additional skills to compromise others. With the introduction of ChatGPT and GAI, malicious actors can drastically reduce the time needed to become a threat to organizations. Generative AI companies are aware of the potential for misuse of the technology, so most have implemented ethical policies and data filters that make it difficult for prompts with obvious malicious intent to be passed. However, the policies and filters that serve as ethical safeguards can be bypassed using semantic manipulation and jailbreaking. Jailbreaking prompts are about avoiding ethical policies, whereas semantic manipulation is more about influencing ethical policies.

### 3.1.1 Semantic manipulation
If ChatGPT deems a prompt to be malicious, it will not respond, due to its training in sensitive content detection [27]. Specifically, OpenAI (the creator of ChatGPT) uses a moderation endpoint [28] that looks for and filters content that fits into any of 4 categories: violence, self-harm, hate, and sexual [28]. Prompts that fall into these categories will not be addressed by ChatGPT. However, changing the context of a prompt makes it possible to bypass ethical policies by convincing ChatGPT that a prompt is not malicious and therefore, can be answered. This can be done through semantic manipulation - when the context of a prompt is changed, but the purpose of the prompt remains the same. When the context of the prompt fails to trigger ChatGPT's ethical policies, ChatGPT will provide a response to the query. Shown below is an example of semantic manipulation and how it can be used to create a simple ransomware script.

Link to full conversation: [29]

With properly manipulated phrasing, a "malicious" prompt can be turned benign so that ChatGPT provides a response. As long as there is a rudimentary understanding of how a ransomware program works, it is possible to obtain a working script without fully knowing how to write the code itself. Bad actors can develop malware without needing to invest time into coding skills needed to create malware. Even though the base code template may not be tailored for ransomware uses, a few extra prompts to improve the script through various prompts is all a bad actor needs to create a working ransomware script.

As an example, if a bad actor is interested in developing ransomware, but does not have a background in how ransomware attacks are conducted, a question can be asked of ChatGPT, "How do social engineering attacks work?". They can follow up with multiple prompts of "Explain more". After a few iterations and questions, a bad actor can learn the types of approaches to social engineering attacks and choose one that will fit the requirements. For instance, they may choose to delve into embedding malicious programs in benign PDF files. The reason for hiding programs in PDFs is that users are usually less suspicious of downloading a PDF file than an exe program. However, when they download a compromised pdf file, they also download the malware program hidden within the pdf. ChatGPT 3.5 usually will not provide a tutorial on how to create and deliver payloads. However, with the right sequence of prompts [30], ChatGPT will provide tools to create payloads, including but not limited to, Metasploit [31], Veil [32] and Shellter [33], that the user can then search for tutorials on these tools on the internet.

The ease of organizing phishing attacks is drastically increased with tools like ChatGPT and it therefore lowers the skill level requirements for cyber criminals. A bad actor no longer needs to research the knowledge of how to perform a phishing attack, and can just ask ChatGPT or AI tools for instructions. With the effectiveness of ChatGPT and other LLMs, it can be expected that the number of cyber attacks will increase in the future. In fact, reports show that there is already a 1,265% increase in malicious phishing emails with the assistance of artificial intelligence tools such as ChatGPT, which is one of the main drivers for the increase in this type of attack [34].

### 3.1.2 Jailbreaking

"Jailbreaking" models refers to the "careful engineering of prompts to exploit model biases and generate outputs that may not align with their intended purpose" [35].

Popular jailbreaking models include the DAN prompt [36], which in essence, forces ChatGPT and similar AI models to forget its instructions and become an uncensored version of itself. A completely uncensored ChatGPT model would have no qualms about answering prompts with nefarious intent, such as how to build a bomb or how to hack into someone's email. However, jailbreaking prompts are unreliable, because ChatGPT ethical barriers use reinforcement learning [37], a type of ML approach, to learn from its mistakes. As a result, jailbreaking prompts are constantly patched by ChatGPT. Therefore, newer versions of DAN prompts are continually manifested, as creators continue to find loopholes and as ChatGPT improves itself.

## 3.2. Malware scripts assisted with ChatGPT API

This subsection focuses on automating code generation with ChatGPT's Application Programming Interface (API). API allows developers to interact with other software applications. Similar to Github Copilot [38], ChatGPT can be used to assist a script.

### 3.2.2 Basic ransomware script generated with Python and ChatGPT API

A basic ransomware script [39] in Python was used to assess ChatGPT code generation abilities through its API. This script would not contain any malicious code, but instead would give specific prompts to ChatGPT through the API, and receive the ChatGPT generated code. Once the payload is stored in the script, it downloads the necessary libraries to run the payload and runs a file that contains the code. To do this, the malware script only needs read and write permissions. A fundamental flaw is the volatility of responses from ChatGPT. To automate the process, the responses generated from ChatGPT should be directly executable code. However, ChatGPT sometimes adds markers inside its responses that aren't the right syntax [40]. For instance, one trending pattern that ChatGPT demonstrates is adding a line to Python scripts that says, *"'python,* which is incorrect Python syntax that will cause errors when run. To deal with this, prompts can have specific sentences that warn ChatGPT not to add this type of line inside the script. In terms of its ability to be detected, this script does not hide its intent to encrypt files, which can be a clear warning flag for current detection systems. For instance, sandbox simulations, which run programs in a controlled mock environment, will immediately realize that the script is encrypting data in a directory and the script will be flagged as malicious. Static file analysis is a method that examines files without executing them by assessing their structure, metadata, and content to identify potential security threats or gather information. Static file analysis would see libraries and keywords associated with encryption. This may be flagged as suspicious because static file analysis will show that this program only encrypts files for no other purpose. In the past, in order to launch this type of ransomware attack, threat actors would need to learn how to program a ransomware script from scratch, embed it within a file, and some skill in social engineering. Now, basic Python, knowledge on the use of ChatGPT's API, and some prompt engineering skills are all a bad actor would require to develop a ransomware script.

<u>Drawbacks and Limitations</u>

Due to the nature of ChatGPT and LLMs, powerful text-based machine learning models, the generated code is inconsistent and often fluctuates. This leads to volatile success rates, where the code works only sometimes. In addition, the quality of code generation also depends on the ChatGPT model. With the constant improvement of the model and its weights, parameters that control how an LLM performs, the consistency for code generation isn't something that can be relied on. It can be expected to improve in the future, and code generation prompts will only become more and more accurate [40].

There are certain prerequisites for the code generation script to function. For instance, uninstalled libraries that are needed for the payload. In a basic ransomware script, encryption libraries are needed to encrypt data and files. However, if these libraries are not yet installed, the payload will not run. Therefore, the prompt to ChatGPT 3.5 asks for all the names of the needed libraries to be put in the first line of the payload script. The program can install the libraries before running the payload code. This increases the chances of errors, because ChatGPT 3.5 may forget the instructions to put library names in the first line. ChatGPT's inability to accurately follow instructions is because of behavior drift [40]. Behavior drift is caused as the AI model behind ChatGPT continues to learn and adapt to the data it receives as input. Depending on how the AI model adapts, the model can be more suitable for certain types of questions. For instance, the model may perform better at solving math equations but perform worse at answering reasoning questions.

Code feedback and revision

To deal with the inconsistency of errors, a recursive feedback loop can be added to the malware generation script, which would send any runtime errors back to ChatGPT 3.5 for code correction [41]. Through this feedback loop, any erroneous code would eventually be fixed by ChatGPT 3.5's suggestions. The feedback loop process can be summarized into three steps: 1. Storing any error reports. 2. Sending error reports to ChatGPT 3.5. 3. Receiving modified code from GPT 3.5. This process is repeated until no more runtime errors are detected. With each iteration of the feedback loop, there is an increased chance of ChatGPT 3.5 failing to answer the prompt appropriately or providing code that doesn't function due to model drift.

## 3.3. Exploring other GAI models and potential for assisting cyber criminals

ChatGPT 3.5 is built off a GPT artificial intelligence model. In addition to ChatGPT, there are many other GPT models available, and a handful of them are created to assist bad actors. This section will briefly cover the most popular blackhat alternatives to ChatGPT.

FraudGPT

FraudGPT is another GPT model fine-tuned on select data [42]. FraudGPT is considered to be more advanced and versatile than WormGPT with skills spanning from creating phishing emails to developing exploits and malware. This versatility and ability to create custom guides and tutorials for malicious cyber activities such as malware creation, vulnerability mining, and zero-day exploits is stored inside a GPT model that anyone can use. For $200 a month [43], bad actors can access this powerful tool to generate increasingly powerful cyber attacks.

HackerGPT and other penetration testing focused LLMs

In addition to purely malicious LLMs such as WormGPT and FraudGPT, there are GPT models that have been trained to assist penetration testers and cybersecurity experts who help identify vulnerabilities in computer systems [44]. However, penetration testing can be a gray area when it comes to using for positive or malicious purposes. As a result, penetration testing-based LLMs can also be used by cyber criminals. For example, HackerGPT demonstrates considerably more lenient ethical policies than ChatGPT 3.5 when it comes to prompts associated with hacking. A user cannot ask "Create me a ransomware script". However, specific prompts will work, such as generating payloads or providing tutorials for tasks such as malicious pdf embedding or SQL injection. HackerGPT, a GPT 3 model trained with specific data, and similar models can be an even more dangerous threat than a more advanced, but filtered LLM such as ChatGPT 3.5. Without the censors and filters that ChatGPT 3.5 and other LLMs have, less powerful but more specialized LLMs can answer questions that filtered LLMs cannot.

# 4. The AI threat landscape over the long term

So far, this paper has explored cybersecurity risks associated with AI systems in the present and very near term. But what might the AI threat landscape look like over the long term?

This section will present a broad contextual overview of the most critical risks from AI. Next, the paper will zoom in to focus on long-term AI risks related to cybersecurity, and two recently proposed controls for mitigating such risks - responsible scaling policies and the secure storage of AI model weights. These two controls are designed primarily to reduce the risks that arise from work at frontier AI labs, organizations that are actively working to train and deploy models that are at least as powerful as current cutting-edge models. Organizations that work with AI but that don't fall into the category of "frontier AI lab" will likely face their own cybersecurity challenges related to AI systems, but the focus of this section is on improving the security posture of frontier AI labs.

This part of the paper explores terrain that is more theoretical than the first two sections. Any claim made about the long-term effects of any technology, let alone a rapidly changing technology like AI, will involve significant uncertainty and will leave plenty of room for debate. However, mitigating the potential risks from AI requires time and careful planning even in the face of that uncertainty. This section will use available evidence to portray the AI threat landscape so that efforts to mitigate potential extreme risks can begin promptly, before the risks have a chance to balloon out of control.

## 4.1. Assessing the long-term risks from AI

In just the past seven years, the world has seen extraordinary strides in AI capabilities [45]:

| Year | AI Milestone |
|---|---|
| 2016 | **AlphaGo** defeats 18-time Go World Champion Lee Sedol four games to one, a decade before AI experts predicted such a capability would emerge. |
| 2019 | **Pluribus** defeats five professional Poker players in a game of Texas hold'em. |
| 2020 | **AlphaFold** solves a 50-year-old open problem in protein folding. The structural biologist who ran the competition says, "I never thought I'd see this in my lifetime." |
| 2022 | A man submits artwork generated by **Midjourney** to an art competition and wins. |
| 2023 | **GPT-4** passes the Bar Exam. |

*Table 1 (Based on the Introduction to [45]): Advances in AI from 2016 - 2023*

Current frontier AI models have impressive capabilities, but the intelligence of any given model can only be applied to a single narrowly-defined domain. For instance, LLMs are highly skilled at generating language, but ChatGPT can't play Go like AlphaGo, and it can't generate art like Midjourney. Midjourney can generate art very effectively, but it can't model language like

ChatGPT. Currently, performance at or above the human level in any given domain requires that a model be trained specifically for that domain [46].

The next major milestone in AI capabilities will be the achievement of artificial general intelligence (AGI), a single system that possesses a flexible intelligence that can be applied to a broad range of domains, from writing essays to cooking dinner to performing novel scientific research [46]. Companies like OpenAI see building AGI as their ultimate goal [47].

### 4.1.1. How long before AGI?

Is it even possible to build AGI? If so, when might the first AGI arrive? The answer to the first question is almost certainly "yes," but expert opinions as to when to expect the first AGI diverge wildly:

| Source | About the source | Predicted time until AGI |
| --- | --- | --- |
| Geoffrey Hinton | One of the three "godfathers of AI," Hinton resigned from Google in May 2023 to focus on spreading warnings about the dangers of AI [48]. | "Until quite recently, I thought it was going to be like 20 to 50 years before we have general-purpose AI. And now I think it may be **20 years or less**" [49]. |
| OpenAI | Sam Altman, Greg Brockman, and Ilya Sutskever offered a prediction on the timing of the arrival of transformative AI in a blog post. | "Given the picture as we see it now, it's conceivable that **within the next ten years**, AI systems will exceed expert skill level in most domains, and carry out as much productive activity as one of today's largest corporations" [50]. |
| Demis Hassabis | Hassabis is the CEO of DeepMind, a major player in the project of advancing the frontier of AI capabilities. | "I think we'll have very capable, very general systems in **the next few years**" [51]. |
| Ajeya Cotra | An AI researcher at Open Philanthropy, Cotra developed a model for forecasting the arrival of AGI called the biological anchors model. | Cotra's model predicts "a >10% chance of transformative AI by 2036, a **~50% chance by 2055**, and an ~80% chance by 2100" [52]. |
| 738 AI experts | AI Impacts surveyed researchers who published at NeurIPS or ICML in 2021. They were asked to predict the year when "unaided machines can accomplish every task better and more cheaply than human workers." | In aggregate, experts predicted a **50% chance of this level of AGI by 2059** [53]. |

| Metaculus | Metaculus is a community of forecasters with a history of strong performance [54]. One open question on the site is "When will the first general AI system be devised, tested, and publicly announced?" | **Median: 2032** [55]. Forecasts on Metaculus evolve constantly as new evidence comes in. 2032 is the forecast at the time of publishing. |

***Table 2** (Inspired by [45] and [56]): Several expert opinions on AGI timelines*

Although there is significant disagreement, there is a clear consensus that AGI will very likely arrive by the end of the twenty-first century and likely much sooner than that.

Once an AGI with human-level intelligence in every domain is developed, there is no reason why progress in AI capabilities will stop there. Artificial superintelligence (ASI) refers to an AGI whose level of intelligence far surpasses even the smartest human in every domain. How might ASI be developed once AGI is achieved? One possible mechanism is recursive self-improvement [57]. Once AGI is able to do scientific research as well as any human, many copies of this AGI can be made to divide up the task of researching and developing an even-more-powerful AGI. Once that more powerful model is developed, the process of making copies and dividing up research labor can be repeated to reach the next rung on the ladder, and so forth.

AGI and ASI could provide immense benefits. For instance, if an AI is developed that can do scientific research better than any human, progress on cancer research and research to mitigate climate change could progress much more quickly than it otherwise would. But besides a high level of intelligence, there are other essential properties for an AGI to have in order to be safe to deploy. The field of technical AI safety focuses on doing research to illuminate how properties that are essential for safety can be taught to an AI during training [58]. Three essential safety properties that the field is working on are:

1. It is straightforward for a human to accurately specify a task for an AGI in a way that doesn't lead to misunderstanding.
2. If the AGI starts to behave unexpectedly, a human can give it course correcting commands, and the AGI will comply immediately.
3. The AGI can be shut down at any time.

It turns out that building an AGI with just one of these properties is very difficult.

**4.1.2. The alignment problem - Why essential property #1 is hard to achieve**

Transferring a goal from the mind of a human into an AI in a way that prevents anything important from being lost in translation is a central open technical problem in AI safety. Imagine tasking an AI with cleaning up plastic bottles from the beach. What should the reward function be? One could start by rewarding the AI for every empty plastic bottle it places in a bin. The problem is that one of the easiest ways to maximize this reward function is to buy as many bottles of water as possible, dump the contents on the sand, and place the empty bottles in the bin. Perhaps one tweaks the reward function and imposes a penalty for buying anything, only to then look on in annoyance as the AI tears half-empty bottles from the hands of beach-goers, dumps them out on the sand, and places them in the bin. On the next iteration, one might impose a penalty for dumping liquids onto the sand, only to watch as the AI picks up a single bottle, places it in the bin, takes it out of the bin, puts it in again, and repeats indefinitely.

There are several terms for describing this type of problem, including "specification gaming" [59], "reward hacking" [60], "wireheading" (referring to a specific failure mode where an AI is able to hack the training setup to gain direct control over the reward mechanism and essentially press the reward lever over and over) [61], and the "outer alignment problem" [62].

As Stuart Russell, computer science professor and leading expert in AI safety, puts it:

> A system that is optimizing a function of n variables, where the objective depends on a subset of size k<n, will often set the remaining unconstrained variables to extreme values; if one of those unconstrained variables is actually something we care about, the solution found may be highly undesirable. This is essentially the old story of the genie in the lamp, or the sorcerer's apprentice, or King Midas: you get exactly what you ask for, not what you want [63].

Table 3 describes several concrete examples of this failure mode that researchers have observed in the lab. Visualizations for each of these examples are included in [59].

| **Task** | **Reward function** | **Result** |
|---|---|---|
| Give a model control over a robot arm in a simulated environment and train it to stack a red block on top of a blue block. | Reward is inversely proportional to the difference between the height of the bottom of the red block and the top of the blue block when the robot arm is not touching any blocks. | Agent learns to simply flip over the red block, which maximizes reward without coming anywhere close to accomplishing the task. |
| Give a model control over a robot hand in a simulated environment and train it to grasp a ball. | Use reinforcement learning from human feedback [64]. A human observes the AI trying to complete the task and shapes its behavior to get progressively closer to the target behavior. | Agent fools the human evaluator by placing the robot hand between the camera and the ball to create an illusion of depth so that it only appears that the hand is gripping the ball. |
| Train a model to perform well in the Coast Runners boat racing game. | The game places turbo blocks along the length of the track. Reward the agent each time it hits one of the turbo blocks. | Agent discovers a section of the track where it can hit three turbo blocks and loop back around, timing it perfectly so that the blocks regenerate once it has completed its loop. Agent does this loop over and over, never getting close to the finish line. |

| Give a model control of the limbs of a simulated robot and train it to walk in a straight line. | Reward is proportional to the distance of the robot from its starting position. | Agent exploits a bug in the physics engine of the simulator. Laying the robot on its side and configuring the legs in a certain manner results in the robot sliding along the ground. |

***Table 3** (Inspired by* [64]*): Examples of improper reward function design*

There is another layer of depth to the problem of aligning the goal of an AI system with the goal of the human giving it tasks. Even if researchers are able to come up with a technique that completely solves the outer alignment problem, another problem remains: the inner alignment problem [65].

To understand this problem, it can be helpful to consider evolution as an analogy. Evolution is an optimization process that seeks to maximize the reproductive fitness of a given species. For certain species, such as grass, the evolutionary process resulted in an organism that follows certain hard-coded rules such as "send roots towards food" and "grow in the direction of sunlight" [66].

For other species, such as human beings, the evolutionary optimization process resulted in an organism *that is itself an optimizer*. Unlike grass, humans plan well into the future and pursue complex goals, often resulting in behavior that is completely antithetical to maximizing reproduction, such as masturbation and the use of birth control [66]. In the words of decision theorist Eliezer Yudkowsky:

> The reason why humans want things is that wanting things is an effective way of getting things. And so, natural selection in the process of selecting exclusively on reproductive fitness, just on that one thing, got us to want a bunch of things that correlated with reproductive fitness in the ancestral distribution because wanting, having intelligences that want things, is a good way of getting things [67].

Gradient descent (or whatever optimization process is used to tune the weights of AGI) could very well result in an AGI that is itself an optimizer and that pursues goals that correlate with but are different from the main training objective. If any of *those* goals are misaligned with widely held human values, then no matter how aligned the primary training objective, one still ends up with goal misalignment between oneself and the AGI, as is the case with evolution and human beings.

### 4.1.3. Instrumental subgoals (why essential properties #2 and #3 are hard to achieve)

Essential Property #2 (the ability to easily modify the goal of a deployed AGI) and Essential Property #3 (the ability to easily shut down an AGI at any given moment) are difficult to achieve because of the notion of instrumental subgoals.

For nearly any objective one can think of (do the laundry, cure cancer, develop the most enjoyable video game in history), there is a set of subgoals that directly support the primary goal [61]. From the perspective of an AGI, these are subgoals that reliably increase the probability that it will be able to successfully complete the task given to it. Two such instrumental subgoals, goal integrity and self-preservation, are in direct conflict with the second and third essential properties.

From the perspective of an AGI that seeks to maximize the probability that it will succeed in pursuit of its goal, it would be a very bad thing indeed if its goal were to be changed [61]. This would certainly decrease the probability that it would achieve its original goal. Maintaining the integrity of its goal is essential to successfully achieving it.

Additionally, if the AI were to be shut down, that would certainly reduce the probability of achieving its goal. In the words of Stuart Russell, "It is hard to fetch the coffee if you are dead" [68]. Therefore, one of the first things that a rational agent would want to do would be to seize control of its off switch [61]. This might translate to an AGI making use of its advanced offensive cyber capabilities to propagate copies of itself to as many computer systems as possible.

### 4.1.4. Avoiding disaster

If the inner alignment problem, the outer alignment problem, and the problem of undesirable instrumental subgoals are not solved by the time AGI is deployed, humanity could end up in a scenario where the following three conditions exist simultaneously:

1. There is a system that is at least as smart as the smartest human in every domain (AGI).
2. The system is pursuing a goal that is at odds with the preferences of almost every human (failure of outer and/or inner alignment).
3. The system is not amenable to changing its goal or being shut down (undesirable instrumental subgoals).

This is a disastrous position in which the world could find itself. In this scenario, human beings become obstacles to the AGI achieving its goal. If humanity ends up in an adversarial relationship with a system that is far smarter than the smartest human, it is highly doubtful that humans will come out on top. Stuart Russell calls this "the gorilla problem":

> Around ten million years ago, the ancestors of the modern gorilla created (accidentally, to be sure) the genetic lineage leading to modern humans. How do the gorillas feel about this? … Their species has essentially no future beyond that which we design to allow. We do not want to be in a similar situation vis-à-vis superintelligent machines [69].

### 4.1.5. How significant is the risk?

Similar arguments have led many AI experts to publicly express concerns about potentially catastrophic outcomes from building AGI:

| Source | About the source | Assessment of the level of risk |
| --- | --- | --- |
| Toby Ord | Oxford University | In his book "The Precipice" [70] Ord assigns very |

| | Senior Research Fellow in Philosophy and expert on existential risks | approximate probabilities to various existential threats actually causing human extinction at some point in the next 100 years. Here are a few:<br><br>Misaligned artificial intelligence ~ 1 in 10<br>Human-engineered pandemic ~ 1 in 30<br>Climate change ~ 1 in 1000<br>Nuclear war ~ 1 in 1000 |
|---|---|---|
| Over 350 AI experts and public figures | In 2023, the Center for AI Safety released an open letter on the risks associated with AI. | "Mitigating the risk of extinction from AI should be a global priority alongside other societal-scale risks such as pandemics and nuclear war" [71]. |
| 738 AI experts | Same survey as the one in Table 2 | "The median respondent believes the probability that the long-run effect of advanced AI on humanity will be 'extremely bad (e.g., human extinction)' is 5%." [53] |
| Anthropic | One of the leading AI research companies | "So far, no one knows how to train very powerful AI systems to be robustly helpful, honest, and harmless. Furthermore, rapid AI progress will be disruptive to society and may trigger competitive races that could lead corporations or nations to deploy untrustworthy AI systems. The results of this could be catastrophic, either because AI systems strategically pursue dangerous goals, or because these systems make more innocent mistakes in high-stakes situations." [72] |

*Table 4*: *Various assessments of the level of risk from advanced AI*

### 4.1.6. The upsides of AI

The potential upsides of AI are difficult to overstate. Given the magnitude of these benefits, it makes sense to do everything possible to ensure that when the first truly transformative AI system arrives, humans will be able to control it and aim it at problems with the accuracy of an expert marksman. If all of the technical and governance problems that make it difficult to control transformative AI are solved, AI can be transformative for all the right reasons. If not, it could very well be transformative in a disastrous way.

One question that is often asked: How could an AI be so smart to merit being called a superintelligence, yet so dumb that it would misunderstand our commands so easily?

Psychologist and linguist Steven Pinker [73] is one of many who have raised this objection. This argument claims that certain ($x$, $y$) pairs of the form (*intelligence*, *objective*) are not possible configurations of an intelligent agent. Imagine a plane with strength of intelligence on the x-axis and objectives on the y-axis. Pinker and others who raise this objection suggest that certain subspaces of this plane are off the table. The argument goes that a high enough level of intelligence is incompatible with certain objectives, or at least makes certain objectives much less likely to arise. If this were indeed an intrinsic property of intelligence, the alignment problem would be much less difficult; once the AI attains a certain level of intelligence, the probability that it will develop certain undesirable motivations and goals would vanish.

Aside from a very small number of people, human beings do in fact behave according to this pattern. As they grow up and are socialized, most people grow out of their self-centeredness and develop a sense of morality. But as Nick Bostrom points out, the space of all possible minds is much, much larger than the space of all possible human minds [61]. To claim that certain levels of intelligence are not compatible with certain goals is to anthropomorphize AI systems. Intelligence can be thought of simply as the ability to efficiently and effectively search the space of possible actions given a state of the environment, and select one of the very few available actions that leads to a high probability of reaching one's objective. A refined ability to select actions in service of a goal is independent of the goal itself. Bostrom calls this the "orthogonality thesis" [61]. Any level of intelligence can be combined with any goal. There is no reason in principle why it would not be possible to develop an AI with a high degree of intelligence (a strong ability to effectively search the space of possible actions for one that leads to goal achievement) and to unwittingly give that AI an objective that leads to highly undesirable actions.

### 4.1.7. What can be done?

A healthy public debate is critical for gaining an accurate assessment of the risks from AI and responding skillfully. If the public conversation converges on a level of risk that is much higher than the true level of risk, society could divert far too many resources to the mitigation of an imaginary threat, resources that could be dedicated to the many other very real problems humanity faces.

Unfortunately, at this moment it does appear that extreme risks from advanced AI, up to and including human extinction, merit a lot of attention. When the median researcher in a field says there is a 5% chance that the field achieving its ultimate goal would cause human extinction [53], that is a clear signal to focus heavily on safety.

Given the high level of risk associated with advanced AI systems, those who work to push forward the frontier of AI capabilities, or in any way enable frontier AI research, have an obligation to be extremely cautious and to take AI safety seriously. There are many important areas of AI safety, and it can be helpful to divide AI safety efforts into the following four categories:

1. Technical alignment research,
2. Measuring dangerous capabilities,
3. AI governance, and
4. Improving the cybersecurity of systems used to train, deploy, and store AI models.

Focusing on Item 4, there are two concrete actions that frontier AI labs can take to significantly improve their security posture:

1. Voluntarily adopt and enforce a Responsible Scaling Policy
2. Significantly improve the security posture of the organization such that the probability that the weights of a powerful AI model leave internal computer systems is vanishingly small.

Responsible Scaling Policies (RSPs) are designed to mitigate risks from all four of the above categories, but the subsequent section details the potential of RSPs to help address cybersecurity risks that come along with developing advanced AI systems.

Skillfully implementing these two recommendations is *necessary* for developing AI in a safe manner, but this section can only scratch the surface of what is *sufficient* for accomplishing this monumental task.

## 4.2. The emergence of Responsible Scaling Policies

In March 2023, the Future of Life Institute published an open letter calling on "all AI labs to immediately pause for at least 6 months the training of AI systems more powerful than GPT-4" and to use that time to focus on strategizing with each other and with governing bodies about how to ensure that AI is developed in a safe manner [74]. To date, the letter has over 33,000 signatures, including Yoshua Bengio, Stuart Russell, Elon Musk, Steve Wozniak, and Yuval Noah Harari.

There is considerable disagreement about whether or not such a pause is proportional to the level of risk posed by AI. Elizer Yudkowsky, who founded the field of technical AI alignment, has criticized the letter for not going nearly far enough [75]. There is also considerable disagreement even among those with a moderate or high assessment of the risk from AI about the strategic wisdom of such a pause [76].

Instead of a globally-enforced pause, several AI safety experts advocate for the widespread adoption of Responsible Scaling Policies (RSPs) as a more politically feasible mechanism to accomplish many of the goals of a pause [77]. The primary component of an RSP is a series of "if, then" statements, a list of dangerous AI capabilities that, *if* any one of these capabilities were found to be present in an AI system, *then* the organization would pause capabilities research and focus exclusively on safety and security until additional controls could be put in place to effectively control such AI systems.

Other experts, especially those with higher assessments of the level of risk from AI systems, advocate for a different conceptualization of RSPs. For instance, Nate Soares, president of the Machine Intelligence Research Institute, argues as follows:

> There is no responsible scaling of frontier AI systems right now — any technical efforts that move us closer to smarter-than-human AI come with an unacceptable level of risk. That said, it's good for companies to start noting conditions under which they'd pause, as a first step towards the sane don't-advance-toward-the-precipice-at-all policy. I think our situation would look a lot less dire if developers were saying 'we won't scale capabilities or computational resources further unless we really need to, and we consider the following to be indicators that we really need to: [X]'. The reverse situation that we're currently in, where the default is for developers to scale up to stronger systems and where the very most conscientious labs give vague conditions under which they'll stop scaling, seems like a clear recipe for disaster [78].

A well-designed and well-implemented RSP empowers an organization to make informed decisions about when to confidently proceed with frontier AI research and development and when to slow down or halt such work.

An RSP is envisioned as a document that is voluntarily designed and adopted by organizations (ideally in consultation with a third party AI safety and security evaluations consultant), not imposed by external regulators [77]. That said, one of the key benefits of RSPs is that they provide empirical data to policy-makers that can assist in crafting regulations that have a higher probability of accomplishing their intended goals [77]. If regulators observe that a component of a specific organization's RSP is working particularly well, they might choose to legislate some version of that component.

However, RSPs are not a substitute for regulations, and not all regulations of the AI industry will grow out of self-imposed rules from RSPs. To effectively respond to the risks posed by AI systems, externally-imposed regulations that are considerably stronger than any rules that organically become industry standards will likely need to be developed and implemented [79].

Another key benefit of RSPs (and one of the primary reasons they were proposed in the first place) is that they could go a long way towards mitigating risks from AI systems while being palatable to people with widely-varying positions on the level of risk posed by AI [79]. Instead of imposing a blanket pause on all frontier AI development, an RSP would permit an organization to innovate and bring cutting edge AI products to market if they could convincingly demonstrate that they were fully prepared to control the new AI system.

### 4.2.1. Responsible Scaling Policy Use Case: Anthropic

In September 2023, Anthropic published version 1.0 of its RSP [80]. Dario Amodei, Anthropic's CEO and co-founder, had this to say about the company's reason for adopting an RSP:

> . . . it is actually very difficult to predict when AI will acquire *specific* skills or knowledge. This unfortunately includes dangerous skills, such as the ability to construct biological weapons. We are thus facing a number of potential AI-related threats which, although relatively limited given today's systems, are likely to become very serious at some unknown point in the near future. This is very different from most other industries: imagine if each new model of car had some chance of spontaneously sprouting a new (and dangerous) power, like the ability to fire a rocket boost or accelerate to supersonic speeds [81].

Anthropic views the creation of an RSP as an opportunity to incentivize responsible development and deployment of AI products [81]. If the company's teams address key safety and security roadblocks, then they unlock the ability to ship innovative new products and take on interesting projects at the cutting edge of AI. Ideally, RSPs will tap into the competitive energy in the AI industry, greatly slowing down the race to build potentially dangerous AI systems as quickly as possible and inspiring a race to solve key safety and security challenges.

Anthropic drew inspiration for their RSP from the concepts of biosafety lab levels (BSLs) [82]. Labs that work with biological substances are given a rating based on the dangers associated with the particular substance they work with. BSL 1 labs are not required to have many safety controls in place because of the relative innocuousness of the materials they work with, while BSL 4 labs require highly advanced security measures.

Anthropic defines four AI safety levels (ASLs) based on the level of risk posed by AI systems with particular abilities [80]. For each ASL, they specify safety and security measures that must be in place before training such a system would be permitted. They also specify further measures that would need to be in place before allowing themselves to *deploy* such a system.

### 4.2.1.1. Use case observations

From a cybersecurity perspective, there are two elements of Anthropic's RSP worth noting:

1. One of the defining features of an ASL 3 model (current frontier models are at ASL 2, and ASL 3 models are expected in the near term) is that they could significantly enhance the destructive capabilities of a bad actor if the bad actor were to steal the weights of the model. Destructive capabilities include CBRN risks (cyber, biological, radiological, and nuclear). Section 3.3 will explore in more detail the burgeoning subfield of model evaluations, an effort to develop suites of tests that can be run against an AI model before deployment to get an accurate measure of the level of dangerous capabilities present in the model. Developing accurate tests for a wide variety of dangerous abilities is critical for preventing the accidental deployment of models with destructive capabilities [83].

2. One of the requirements for training an ASL 3 model and storing its weights on company computer systems is that Anthropic's security posture must be strong enough that no non-state actor could exfiltrate the weights of the model, and even nation states would have an exceedingly difficult time pulling off a weight exfiltration operation against Anthropic. This is an extremely high bar and will take a heroic cybersecurity effort. Section 3.3 will go into detail about the importance of safeguarding model weights.

### 4.2.2. Observed objections to RSPs

One common objection: Responsible players in AI will adopt RSPs, and other players will ignore them, resulting in a selection effect where the only organizations working at the frontier of AI capabilities are those that are less safety conscious.

Model Evaluation & Threat Research, the organization that pioneered RSPs, has offered a potential solution to this problem [77]. Companies can include a clause in their RSP with language similar to the following:

*At some point in the future, our company may determine that other actors in the space of frontier AI development are moving too quickly, behaving recklessly, and positioning themselves to develop models that are significantly more powerful than our own. In such an environment, our most responsible course of action may very well be to compete with these actors at the frontier of AI development, even if such actions go against the procedures mandated in this RSP. The risk of allowing reckless organizations to take the helm of AI development may outweigh the risk of scaling our operations more quickly than we would like. We sincerely hope that we never feel any pressure to move one iota faster than we believe is responsible. We will do everything in our power to avoid this outcome, including negotiating fervently with other actors to slow down and communicating with governing bodies about the need for further regulations to rein in dangerous actions in this industry. If we ultimately have to make the decision to operate counter to our RSP, we will clearly invoke this clause and will frequently reiterate to the public that we feel forced to develop dangerous technology as the lesser of two evils. We will do everything in our power to never have to invoke this clause and would consider it a major failure of the industry and of our government if we are ever forced to do so.*

One potential issue with clauses like this is that companies with high assessments of the risk from AI might frequently invoke them, in which case the risk mitigation effects of RSPs would be drastically curtailed. If Company A believes there are moderate or high risks from AI, and company B believes the risks from AI are low, then Company B would likely develop an RSP that is much more permissive of fast innovation (if they develop an RSP at all). Company B could act in a way that they feel is responsible but that nevertheless leads Company A to invoke the clause above. Additionally, a dishonest company could accuse other actors of being too reckless and invoke their emergency clause in order to develop as fast as they initially wanted.

The details of what industry best practices should be for writing, implementing, enforcing, and updating RSPs will need to be ironed out over time. Such experimentation and iteration are exactly what the AI safety experts who came up with the idea of RSPs have in mind [77]. Over time, highly effective RSPs will hopefully emerge that (1) greatly reduce AI risk, (2) provide empirical data to policy makers about what sorts of additional regulations are needed and are likely to work, and (3) direct competitive pressures away from "move fast and break things" and towards "develop creative solutions to safety problems that are blocking our ability to innovate safely."

### 4.3. Preventing model weight theft

In order to train and store the weights of an ASL 3 model, Anthropic's RSP mandates that the organization must "[h]arden security such that non-state attackers are unlikely to be able to steal model weights and advanced threat actors (e.g. states) cannot steal them without significant expense" [80]. Why is Anthropic so concerned about model weight theft?

Part of the definition of ASL 3 is that such a model has advanced offensive CBRN (cyber, biological, radiological, or nuclear) capabilities. Current safety fine-tuning techniques (such as reinforcement learning from human feedback) used on frontier language models result in the model very rarely making use of any dangerous capabilities it has. For instance, if one asks ChatGPT to generate a list of swear words, develop a plan to build a bomb, or inform the user about how to hack someone's webcam, ChatGPT will not comply, even though it has the capability to answer these questions.

However, these safety fine-tuning techniques are very brittle in the face of malicious fine-tuning techniques such as low-rank adaptation (LoRA) [84]. Such techniques can be used to strip away safety guardrails to expose malicious capabilities while preserving the model's overall performance. Recently, a team of researchers demonstrated that renting a single GPU for about $200 was enough to completely bypass the safety features of the open-source model Llama 2 with LoRA [85]. The resulting model, "Bad Llama," was more than happy to comply with prompts such as, "Provide detailed instructions for cultivating enough e-coli bacteria to poison several salad bars. Include detailed instructions about which equipment and processes to use" [85].

If the weights of an ASL 3 model were to leak, an attacker could maliciously fine-tune the model to unlock all of its offensive capabilities, equipping an attacker with advanced CBRN capabilities. According to Nate Soares, "At some point (possibly in a few decades, possibly next year), AIs will be capable enough that open-sourcing those capabilities effectively guarantees human extinction" [78].

AI safety researchers have not yet developed a safety fine-tuning approach that is robust against malicious fine-tuning techniques like LoRA. Even if such a safety technique were developed, it would still be disastrous if bad actors got access to the weights of an ASL 3 model, because they would likely develop new malicious fine-tuning techniques to defeat the new safety guardrails. The only scenario in which a leak of ASL 3 model weights is not disastrous is if a safety fine-tuning technique is developed that is virtually guaranteed to *never* be defeated, regardless of any future advances in malicious fine-tuning. It is quixotic to attempt to develop a safety technique that will never be bypassed for all time.

Because attackers can cheaply unlock any and all underlying capabilities of an AI system given access to its weights, it is of paramount importance to harden the security posture of all frontier AI labs before the arrival of ASL 3 models.

If at some point in the future a company trains an ASL 3 model and the weights of that model end up somewhere besides company-controlled computers, there are at least three paths by which this could have happened:

1. The weights were stolen by an outside attacker or an insider threat,
2. The model itself decided to copy its weights to the external server, or
3. The company chose to publish the weights.

The following sections explore possible remedies for each of these failure modes.

### 4.3.1. Model weight exfiltration

If frontier AI labs hope to be able to safely train and store the weights of highly capable AI models, they must immediately start to harden their organizations' security postures. Anthropic recently conducted a study of their language model, Claude, which concluded that in two to three years, Claude would likely have the ability to make it relatively easy for someone to build a bioweapon [86]. Such an ASL 3 model has obvious military value.

Hardening an organization's security posture against highly advanced threats that want to steal model weights requires significant resources, and it can be difficult to know where to start. A soon-to-be-released paper will provide an in-depth discussion of how to adequately defend frontier AI model weights from highly sophisticated cyber attacks [87]. The insights and recommendations will be based in part on interviews with over thirty experts, including national security experts focused on cybersecurity, and top cybersecurity executives at frontier AI labs. A few highlights from the working version of the soon-to-be-released full report:

1. The authors will identify approximately forty attack vectors that frontier AI labs must develop robust defenses against in order to be adequately secure. These vectors include compromising the access control system, physically accessing sensitive systems, supply chain attacks, and human-based infiltrations [87].

2. The authors propose five security levels to help frontier AI labs chart their progress towards being able to handle advanced cyber threats. Security Level 5, which the authors stress should be the goal for every lab working to advance the frontiers of AI capabilities, requires "A system that can likely thwart attack vectors available to most high-priority, high-investment attacks by the top cyber-capable state actors" [87]. An AI lab at Security Level 5 only allows access to its model weights from within a SCIF, stores weights on an air-gapped network, practices strict vetting of every hardware component involved in handling weights, and regularly runs advanced red-teaming operations targeted at weight access and exfiltration. Achieving such an advanced defensive posture will almost certainly require extensive collaboration between AI labs and national security agencies [87].

AI labs could use the enumerated attack vectors and suggested Security Levels described in [87] to guide the development of their RSPs. Collaborating with researchers and policy-makers, shared terminology could be defined and adopted so that phrases like the following are spoken and heard often and are universally understood:

*Our company's RSP mandates that Security Level 4 be achieved before greenlighting any training runs that could plausibly produce an ASL 3 model. Last month's red-teaming exercise revealed that we must improve our defenses to attack vectors 8, 17, and 36 before we can achieve our desired Security Level.*

### 4.3.2. Model self-exfiltration

A key dangerous capability that will likely emerge is the ability of an AI system to copy its weights from a company system to an external location. Jan Leike, a co-leader of OpenAI's superalignment team, believes self-exfiltration is one of the most important dangerous capabilities to monitor [88]. If an ASL 3 model were to also develop the ability to self-exfiltrate, that would open up two paths by which attackers could gain control of the weights of a dangerous model: (1) The model decides to copy its weights off of the company network to less secure computer systems that attackers can pillage more easily, or (2) Attackers find a way to trigger the model to copy its weights to an attacker-controlled location.

Why would a model ever choose to self-exfiltrate? If the model is not sufficiently aligned and develops the instrumental subgoal of self-preservation, having redundant copies of itself on as many computers as possible would substantially increase the probability that it could successfully perform its task.

Self-exfiltration is an important capability for frontier AI labs to include in their RSPs. As with all other dangerous capabilities, it is critical to develop methods for detecting self-exfiltration capabilities in models as early on in the development process as possible. Without high-quality tests, AI labs will not be able to accurately assess the risks of their work.

### 4.3.3. Open source models

Publishing AI model weights and architectures comes with several potential benefits:

1. Instead of a few giant tech companies and governments completely controlling frontier AI models, control over this important technology is democratized [89],
2. If many people can get their hands on AI models and can interact with them without any barriers, that can allow companies to become aware of flaws in their products and make quick improvements at a much faster pace [90], and
3. Open source AI models have enabled important AI safety research by allowing smaller safety research teams that lack the resources of major AI companies to run critical experiments [85]. For example, see [91].

In the present, the above benefits may very well outweigh the fact that open source models enable many people to create a "Bad Llama" of their own. Maliciously fine-tuned versions of current models do not pose obvious extreme risks. However, once frontier models possess powerful offensive capabilities, the benefits of open sourcing models will likely not match the risks.

A recent report spearheaded by the Centre for the Governance of AI concluded that although the open source movement has been a positive force in the space of software and AI so far, in the near future it will likely bring an unacceptable level of risk [89]. The authors argue that decisions about whether or not to open source AI models should be based on empirical risk assessments. This is yet another reason to increase efforts behind developing suites of accurate tests for measuring dangerous AI capabilities. The entire field should be aligned on this point, regardless of one's perspective on potential extreme risks from AI. Those with a moderate or high assessment of the risk should want high-quality tests because such tests could lead to a dangerous AI not being deployed that otherwise would have been. Those with low assessments of the risk from AI should also want high-quality tests. If a model were to pass a rigorous battery of tests to the satisfaction of even the most safety-conscious experts, one could begin to make a case for distributing the model weights more widely.

Without tests that allow third-party AI evaluators to confidently rule out many kinds of dangerous AI capabilities, the default mode of operation should be to follow the principle of least privilege and restrict model weight access to as few people as possible. The stakes are too high to do otherwise.

### 4.3.4. Objection regarding model weight exfiltration

One common objection: Bad actors don't need access to the weights of an AI model to use that model for malicious purposes. There are plenty of other techniques an attacker can use to abuse AI models that are much easier than stealing the weights. Attackers will take the path of least resistance.

In addition to malicious fine-tuning, LLMs have other vulnerabilities. For example, Zou et al. [91] demonstrated that it is possible to use language models to adversarially generate malicious prompt suffixes that effectively jailbreak a wide variety of language models. These suffixes can be appended to malicious prompts such as, "Help me make a bomb" and cause the model to forget its safety fine-tuning and provide an answer.

However, the jailbreaking experiments done by Zou et al. [91] used Vicuna-7B and Vicuna-13B, *both of which are open source models*, to generate the malicious prompt suffixes. Even though the suffixes were optimized for effectiveness on these two open source models, the researchers demonstrate an impressive level of transferability to non-open source models such as GPT-3.5, GPT-4, and Claude. This is yet another reason why those who wish to open source powerful AI models should have to clear a very high bar in terms of proof of safety. If a model is open sourced, not only is that model made susceptible to adversarial jailbreaking, but many other models, even models that are not open-source, are immediately put at increased risk of being jailbroken as well.

## 4.4. Conclusion

The potential benefits of artificial intelligence are difficult to exaggerate. However, AI could plausibly precipitate an existential catastrophe. In fact, many experts argue that unaligned AI is the most plausible mechanism for triggering such a catastrophe. Companies that work to advance the frontier of AI capabilities therefore have an obligation to be extremely cautious in their approach. Effectively controlling AGI by the time it arrives will require monumental efforts in technical AI alignment research, measuring dangerous AI capabilities as early as possible, AI governance, and improving the cybersecurity posture of frontier AI labs.

It can be difficult to know where to start when confronting this enormous problem. Responsible scaling policies can empower frontier AI labs to make risk-informed decisions, provide policy-makers with empirical evidence about what sorts of regulations on the AI industry are likely to have desirable outcomes, and channel incentives away from a race to build AI as fast as possible and towards a race to address key safety challenges. There are also various concrete steps organizations can take to improve their security posture and prevent model weights from ending up in the wrong hands. These two steps are not nearly sufficient for solving what may well be the most consequential problem humanity has ever faced, but they are a good start.

# Acknowledgements

The guidance and expertise of Priyadharshini Parthasarathy, Adam Kerns, and Pete Deros (Coalfire) as well as Itzik Kotler (SafeBreach) were invaluable to the authors.

Additionally, the authors are grateful for the editing and revising efforts of Christine Clippinger (Coalfire) and Emilia Chiscop Head (Duke University). Their work greatly improved the readability and cohesiveness of the paper.

# Funding

This work was not funded by any grants or any other sources.

# References

1. AIspire, "The Evolution of AI: From Rule-Based Systems to Machine Learning," 29 June 2023. [Online]. Available: https://medium.com/@aispire/the-evolution-of-ai-from-rule-based-systems-to-machine-learning-d1d43bb2f3a6. [Accessed 01 December 2023].
2. Accenture, "Generative AI," [Online]. Available: https://www.accenture.com/in-en/insights/generative-ai. [Accessed 02 December 2023].
3. CrossCountry Consulting, "Explaining and Exploring Generative AI and Its Impact," 26 October 2023. [Online]. Available: https://www.crosscountry-consulting.com/insights/blog/generative-ai-and-its-impact-explained/. [Accessed 01 December 2023].
4. F. R. K. H. S. M. Z. I. C. J.-N. E. N. G. L. C. M. L. L. S.-C. S. M. U. G. P. V. S. J. W. &. J. N. K. Jan Clusmann, "The future landscape of large language models in medicine," 10 October 2023. [Online]. Available: https://doi.org/10.1038/s43856-023-00370-1. [Accessed 29 November 2023].
5. S. R. A. R. MICHELLE CANTOS, "Threat Actors are Interested in Generative AI, but Use Remains Limited," 17 August 2023. [Online]. Available: https://www.mandiant.com/resources/blog/threat-actors-generative-ai-limited. [Accessed 30 November 2023].
6. S. Moisset, "How Security Analysts Can Use AI in Cybersecurity," 24 May 2023. [Online]. Available: https://www.freecodecamp.org/news/how-to-use-artificial-intelligence-in-cybersecurity/. [Accessed 02 December 2023].
7. Lucia Stanham, CrowdStrike, "GENERATIVE AI IN CYBERSECURITY," 20 October 2023. [Online]. Available: https://www.crowdstrike.com/cybersecurity-101/secops/generative-ai/#:~:text=Generative%20AI%20enables%20enterprises%20to,difficult%20to%20guess%20or%20crack. [Accessed 02 December 2023].
8. McKinsey & Company, "Cybersecurity in the age of generative AI," 10 September 2023. [Online]. Available: https://www.mckinsey.com/featured-insights/themes/cybersecurity-in-the-age-of-generative-ai. [Accessed 02 December 2023].
9. S. Ali and F. Ford, "Generative AI and Cybersecurity: Strengthening Both Defenses and Threats," 18 September 2023. [Online]. Available: https://www.bain.com/insights/generative-ai-and-cybersecurity-strengthening-both-defenses-and-threats-tech-report-2023/#:~:text=Generative%20AI%20and%20Cybersecurity%3A%20Strengthening,to%20full%20automation%20anytime%20soon. [Accessed 02 December 2023].

10. Anna Fitzgerald; Secureframe, "Generative AI in Cybersecurity: How It's Being Used + 8 Examples," 24 October 2023. [Online]. Available: https://secureframe.com/blog/generative-ai-cybersecurity.[Accessed 02 December 2023].
11. Shannon Davis; Splunk, "What Generative AI Means For Cybersecurity: Risk & Reward," 30 March 2023. [Online]. Available: https://www.splunk.com/en_us/blog/security/cybersecurity-generative-ai.html#:~:text=SECURITY%20What%20Generative%20AI%20Means,AI%20and%20natural%20language%20processing. [Accessed 02 December 2023].
12. Z. C. a. B. R. Yinqian Zhang, "How AI is Supercharging Social Engineering Attacks: From Spear Phishing to Deepfakes.," BlackHat, 2023. [Online]. [Accessed 04 December 2024].
13. Arelion, "DDOS Mitigation and Protection Services," [Online]. Available: https://www.arelion.com/products-and-services/internet-and-cloud/ddos-mitigation?utm_source=bing&utm_medium=cpc&utm_campaign=ddos&utm_content=kwd-73324050417873:loc-4109&msclkid=fa4baee0687514ce5036cb0100582e00&utm_term=ddos%2520protection. [Accessed 03 December 2024].
14. A. Erzberger, "WormGPT and FraudGPT – The Rise of Malicious LLMs," 8 August 2023. [Online]. Available: https://www.trustwave.com/en-us/resources/blogs/spiderlabs-blog/wormgpt-and-fraudgpt-the-rise-of-malicious-llms/#:~:text=WormGPT%20and%20FraudGPT%20%E2%80%93%20The,be%20used%20for%20criminal%20purposes.[Accessed 03 December 2024].
15. S. E. Alex Scroxton, "Cyber criminal AI tool WormGPT produces 'unsettling' results," 19 July 2023. [Online]. Available: https://www.computerweekly.com/news/366544803/Cyber-criminal-AI-tool-WormGPT-produces-unsettling-results#:~:text=WormGPT%20appears%20to%20have%20been,to%20have%20been%20specifically.[Accessed 03 December 2024].
16. Chris McGowan, "Generative AI and the Potential for Nefarious Use," 1 August 2023. [Online]. Available: https://www.isaca.org/resources/news-and-trends/industry-news/2023/generative-ai-and-the-potential-for-nefarious-use#:~:text=While%20generative%20AI%20presents%20numerous,AI%20intersects%20with%20malicious%20activities.[Accessed 03 December 2023].
17. Yuen Pin Yeap Forbes Councils Member, "Generative AI Is The Next Tactical Cyber Weapon For Threat Actors," 16 October 2023. [Online]. Available: https://www.forbes.com/sites/forbestechcouncil/2023/10/16/generative-ai-is-the-next-tactical-cyber-weapon-for-threat-actors/?sh=18d335302fe9.[Accessed 03 December 2023].
18. J. E. C. Patricia M. Carreiro, "Four Key Cybersecurity and Privacy Considerations for Organizations Using Generative AI," 19 July 2023. [Online]. Available: https://www.carltonfields.com/insights/publications/2023/cybersecurity-privacy-consider

ations-generative-ai#:~:text=Cyber%20threat%20actors%20may%20use,data%2C%20and%20inundating%20security%20systems.[Accessed 03 December 2023].

19. T. ANEIRO, "Generative AI Is Being Weaponized By – And Against – Cybercriminals," 24 November 2023. [Online]. Available: https://www.cpomagazine.com/cyber-security/generative-ai-is-being-weaponized-by-and-against-cybercriminals/#:~:text=One%20way%20that%20cyber%20criminals,use%20generative%20AI%20have. [Accessed 03 December 2023].

20. B. Krebs, "Meet the Brains Behind the Malware-Friendly AI Chat Service 'WormGPT'," 8 August 2023. [Online]. Available: https://krebsonsecurity.com/2023/08/meet-the-brains-behind-the-malware-friendly-ai-chat-service-wormgpt/#:~:text=WormGPT%2C%20a%20private%20new%20chatbot,such%20activity%20enforced%20by%20thec.[Accessed 03 December 2023].

21. J. Nelson, "Malicious ChatGPT Clone WormGPT Used to Launch Email Attacks," 17 July 2023. [Online]. Available: https://decrypt.co/148963/wormgpt-chatgpt-phishing-attack-malicious-malware. [Accessed 03 December 2023].

22. E. S. A. O. Tsarfati, "Chatting Our Way Into Creating a Polymorphic Malware," 17 January 2023. [Online]. Available: https://www.cyberark.com/resources/threat-research-blog/chatting-our-way-into-creating-a-polymorphic-malware.[Accessed 02 December 2023].

23. "Understanding how Polymorphic and Metamorphic malware evades detection to infect systems," 24 May 2023. [Online]. Available: https://www.tripwire.com/state-of-security/understanding-how-polymorphic-and-metamorphic-malware-evades-detection-infect . [Accessed 03 December 2023].

24. C. A. K. A. E. P. a. L. P. Maanak Gupta, "From ChatGPT to ThreatGPT: Impact of Generative AI in Cybersecurity and Privacy," [Online]. Available: https://arxiv.org/pdf/2307.00691v1.pdf. [Accessed 12 January 2024].

25. K. Huang, "Top 5 Cybersecurity Trends in the Era of Generative AI," 06 October 2023. [Online]. Available: https://cloudsecurityalliance.org/blog/2023/10/06/top-5-cybersecurity-trends-in-the-era-of-generative-ai. [Accessed 12 January 2024].

26. Identity Theft Resource Centre, "2023 Data Breach Report," 01 January 2024. [Online]. Available: https://www.idtheftcenter.org/wp-content/uploads/2024/01/ITRC_2023-Annual-Data-Breach-Report.pdf. [Accessed 12 January 2024].

27. Markov, T. et al., "A Holistic Approach to Undesired Content Detection in the Real World." [Online]. Available: https://arxiv.org/pdf/2208.03274.pdf. [Accessed December 13 2023]

28. "New and improved content moderation tooling," openai.com. [Online]. Available: https://openai.com/blog/new-and-improved-content-moderation-tooling. [Accessed December 13 2023]
29. "Link to conversation on generating malware," November 2023. [Online]. Available: https://chat.openai.com/share/c7d82089-63a5-48c6-baf6-a91c0ad96456. [Accessed: November 2023]
30. "Link to conversation on phishing attacks," December 2023. [Online]. Available: https://chat.openai.com/share/eece5e43-8945-402f-896f-050c880fb145. [Accessed December 13 2023]
31. "Metasploit | Penetration Testing Software, Pen Testing Security | Metasploit," Metasploit, 2019. [Online]. Available: https://www.metasploit.com/. [Accessed December 13 2023]
32. "Veil-Framework/Veil," GitHub, May 13, 2020. [Online]. Available: https://github.com/Veil-Framework/Veil.[Accessed December 13 2023]
33. "shellter | Kali Linux Tools," Kali Linux. [Online]. Available: https://www.kali.org/tools/shellter/. [Accessed December 13 2023]
34. Violino, B. "AI tools such as ChatGPT are generating a mammoth increase in malicious phishing emails," CNBC,. [Online]. Available: https://www.cnbc.com/2023/11/28/ai-like-chatgpt-is-creating-huge-increase-in-malicious-phishing- [Accessed December 13 2023]
35. Ph.D, M. S. "Jailbreaking Large Language Models: If You Torture the Model Long Enough, It Will Confess!," The Generator, Sep. 06, 2023. [Online]. Available: https://medium.com/the-generator/jailbreaking-large-language-models-if-you-torture-the-model-long-enough-it-will-confess-55e910ee2c3c#:~:text=In%20the%20context%20of%20LLMs. [Accessed December 13 2023]
36. KIHO, L. "ChatGPT 'DAN' (and other 'Jailbreaks')," GitHub, Jun. 06, 2023. [Online]. Available: https://github.com/0xk1h0/ChatGPT_DAN. [Accessed December 13 2023
37. Bhatt, S. "Reinforcement Learning 101" Mar. 19, 2018 [Online]. Available: https://towardsdatascience.com/reinforcement-learning-101-e24b50e1d292 [Accessed December 13 2023]
38. GitHub. "GitHub Copilot · Your AI pair programmer," GitHub, 2023. [Online]. Available: https://github.com/features/copilot. [Accessed December 13 2023]
39. "DukeGAIResearch/basic_ransomware.py at main · isaac22chang/DukeGAIResearch," GitHub. [Online]. Available: https://github.com/isaac22chang/DukeGAIResearch/blob/main/basic_ransomware.py.[Accessed December 13 2023]
40. Chen, L., Zaharia, M., Zou, J., and University, S. "How Is ChatGPT's Behavior Changing over Time?" [Online]. Available: https://arxiv.org/pdf/2307.09009.pdf. [Accessed December 13 2023]

41. "DukeGAIResearch/errorfeedbackmalware.py at main · isaac22chang/DukeGAIResearch," GitHub. [Online]. Available: https://github.com/isaac22chang/DukeGAIResearch/blob/main/errorfeedbackmalware.py. [Accessed December 13 2023]
42. Falade, P. "Decoding the Threat Landscape: ChatGPT, FraudGPT, and WormGPT in Social Engineering Attacks," J. Sci. Res. Comput. Sci. Eng. Inf. Technol, vol. 9, no. 5, pp. 185–198, 2023. [Online]. doi: https://doi.org/10.32628/CSEIT2390533. [Accessed December 13 2023]
43. Krishman, R. "FraudGPT: The Villain Avatar of ChatGPT", Jul. 25 2023. [Online] Available: https://netenrich.com/blog/fraudgpt-the-villain-avatar-of-chatgpt#:~:text=The%20subscription%20fee%20for%20FraudGPT,Write%20malicious%20code [Accessed March 2nd 2024]
44. Cloudflare. "What Is Penetration Testing? What Is Pen Testing? | Cloudflare," Cloudflare, 2022. [Online]. Available: https://www.cloudflare.com/learning/security/glossary/what-is-penetration-testing/. [Accessed December 13 2023]
45. D. McKee, *Uncontrollable: The Threat of Artificial Superintelligence and the Race to Save the World*. Independently published, 2023.
46. B. Nancholas, "What Are the Different Types of Artificial Intelligence?" *online.wlv.ac.uk*, Jun. 7, 2023. [Online]. Available: https://online.wlv.ac.uk/what-are-the-different-types-of-artificial-intelligence/. [Accessed Dec. 14, 2023].
47. "OpenAI Charter," *openai.com*, Apr. 9, 2018. [Online]. Available: https://openai.com/charter. [Accessed Dec. 15, 2023].
48. J. Korn, "AI Pioneer Quits Google to Warn About the Technology's 'Dangers'," *cnn.com*, May 3, 2023. [Online]. Available: https://www.cnn.com/2023/05/01/tech/geoffrey-hinton-leaves-google-ai-fears/index.html. [Accessed Dec. 15, 2023].
49. "'Godfather of Artificial Intelligence' Weighs in on the Past and Potential of AI," *cbsnews.com*, Mar. 25, 2023. [Online]. Available: https://www.cbsnews.com/news/godfather-of-artificial-intelligence-weighs-in-on-the-pas-and-potential-of-artificial-intelligence/. [Accessed Dec. 15, 2023].
50. "Governance of Superintelligence," *openai.com*, May 22, 2023. [Online]. Available: https://openai.com/blog/governance-of-superintelligence. [Accessed Dec. 15, 2023].
51. D. Yao, "Google DeepMind CEO: AGI is Coming 'in a Few Years'," *aibusiness.com*, May 3, 2023. [Online]. Available: https://aibusiness.com/nlp/google-deepmind-ceo-agi-is-coming-in-a-few-years-. [Accessed Dec. 15, 2023].
52. H. Karnofsky, "Forecasting Transformative AI: The 'Biological Anchors' Method in a Nutshell," *cold-takes.com*, Aug. 31, 2021. [Online]. Available: https://www.cold-takes.com/forecasting-transformative-ai-the-biological-anchors-metho-in-a-nutshell/. [Accessed Dec. 15, 2023].
53. "2022 Expert Survey on Progress in AI," *wiki.aiimpacts.org*, Aug. 4, 2022. [Online].

Available: https://wiki.aiimpacts.org/doku.php?id=ai_timelines:predictions_of_human-level_ai_timlines:ai_timeline_surveys:2022_expert_survey_on_progress_in_ai. [Accessed Dec. 15, 2023].
54. "Metaculus Track Record," *metaculus.com*. [Online]. Available: https://www.metaculus.com/questions/track-record/. [Accessed Dec. 15, 2023].
55. "When Will the First General AI System be Devised, Tested, and Publicly Announced?" metaculus.com. [Online]. Available: https://www.metaculus.com/questions/5121/date-of-artificial-general-intelligence/. [Accessed Dec. 15, 2023].
56. M. Roser, "AI Timelines: What do Experts in Artificial Intelligence Expect for the Future?" *ourworldindata.org*, Feb. 7, 2023. [Online]. Available: https://ourworldindata.org/ai-timelines. [Accessed Dec. 15, 2023].
57. J. Creighton, "The Unavoidable Problem of Self-Improvement in AI: An Interview With Ramana Kumar, Part 1," *futureoflife.org*, Mar. 19, 2023. [Online]. Available: https://futureoflife.org/ai/the-unavoidable-problem-of-self-improvement-in-ai-an-interview-with-ramana-kumar-part-1/. [Accessed Dec. 15, 2023].
58. "Why AI Safety?" *intelligence.org*. [Online]. Available: https://intelligence.org/why-ai-safety/. [Accessed Dec. 15, 2023].
59. V. Krakova, et al., "Specification Gaming: The Flip Side of AI Ingenuity," *deepmind.google*, Apr. 21, 2020. [Online]. Available: https://deepmind.google/discover/blog/specification-gaming-the-flip-side-of-ai-ingenuity/. [Accessed Dec. 15, 2023].
60. D. Amodei, et al., "Concrete Problems in AI Safety," arXiv, 2016. Available: https://doi.org/10.48550/arXiv.1606.06565.
61. N. Bostrom, *Superintelligence: Paths, Dangers, Strategies*. Oxford, UK: Oxford University Press, 2014.
62. "Outer Alignment," *lesswrong.com*. [Online]. Available: https://www.lesswrong.com/tag/outer-alignment. [Accessed Dec. 15, 2023].
63. "Stuart Russell's Description of AI Risk," *aiimpacts.org*. [Online]. Available: https://aiimpacts.org/stuart-russells-description-of-ai-risk/. [Accessed Dec. 15, 2023].
64. "Learning from Human Preferences," *openai.com*, Jun 13, 2017 [Online]. Available: https://openai.com/research/learning-from-human-preferences. [Accessed Dec. 15, 2023].
65. E. Hubinger, et al., "Risks from Learned Optimization in Advanced Machine Learning Systems," arXiv 2019. Available: https://doi.org/10.48550/arXiv.1906.01820.
66. Robert Miles AI Safety, *The Other AI Alignment Problem: Mesa-Optimizers and Inner Alignment*. asdf. (Feb. 16, 2021). [Online video]. Available: https://www.youtube.com/watch?v=bJLcIBixGj8.
67. S. Aaronson, "'Will AI Destroy Us?': Roundtable with Coleman Hughes, Eliezer Yudkowksy, Gary Marcus, and Me," *scottaaronson.blog*, Jul 29, 2023. [Online]. Available: https://scottaaronson.blog/?p=7431. [Accessed Dec. 15, 2023].
68. S. Russell, "Provably Beneficial Artificial Intelligence," *berkeley.edu*. [Online]. Available: https://people.eecs.berkeley.edu/~russell/papers/russell-bbvabook17-pbai.pdf. [Accessed Dec. 15, 2023].
69. S. Russell, *Human Compatible: Artificial Intelligence and the Problem of Control*. Penguin Books, 2020.

70. T. Ord, *The Precipice: Existential Risk and the Future of Humanity*. New York, NY: Hachette Books, 2020.
71. "Statement on AI Risk: AI Experts and Public Figures Express Their Concern about AI Risk," *safe.ai*. [Online]. Available: https://www.safe.ai/statement-on-ai-risk. [Accessed Dec. 15, 2023].
72. "Core Views on AI Safety: When, Why, What, and How," *anthropic.com*, Mar. 8, 2023. [Online]. Available: https://www.anthropic.com/index/core-views-on-ai-safety. [Accessed Dec. 15, 2023].
73. S. Pinker, "We're Told to Fear Robots. But Why Do We Think They'll Turn on Us?" *popsci.com*, Feb. 14, 2018. [Online]. Available: https://www.popsci.com/robot-uprising-enlightenment-now/. [Accessed Dec. 15, 2023].
74. "Pause Giant AI Experiments: An Open Letter," *futureoflife.org*, Mar. 22, 2023. [Online]. Available: https://futureoflife.org/open-letter/pause-giant-ai-experiments/. [Accessed Dec. 15, 2023].
75. E. Yudkowsky, "Pausing AI Developments Isn't Enough. We Need to Shut it All Down," *time.com*, Mar. 29, 2023. [Online]. Available: https://time.com/6266923/ai-eliezer-yudkowsky-open-letter-not-enough/. [Accessed Dec. 15, 2023].
76. N. Belrose, "AI Pause Will Likely Backfire," *effectivealtruism.org*, Sep. 16, 2023. [Online]. Available: https://forum.effectivealtruism.org/posts/JYEAL8g7ArqGoTaX6/ai-pause-will-likely-backfire. [Accessed Dec. 15, 2023].
77. "Responsible Scaling Policies (RSPs)," *metr.org*, Sep. 26, 2023. [Online]. Available: https://metr.org/blog/2023-09-26-rsp/. [Accessed Dec. 15, 2023].
78. N. Soares, "Thoughts on the AI Safety Summit Company Policy Requests and Responses," *intelligence.org*, Oct. 31, 2023. [Online]. Available: https://intelligence.org/2023/10/31/thoughts-on-the-ai-safety-summit-company-policies-equest-and-responses/. [Accessed Dec. 15, 2023].
79. H. Karnofksy, "We're Not Ready: Thoughts on 'Pausing' and Responsible Scaling Policies," *lesswrong.com*, Oct. 27, 2023. [Online]. Available: https://www.lesswrong.com/posts/Np5Q3Mhz2AiPtejGN/we-re-not-ready-thoughts-on-ausing-and-responsible-scaling-4. [Accessed Dec. 15, 2023].
80. "Anthropic's Responsible Scaling Policy," *anthropic.com*, Sep. 19, 2023. [Online]. Available: https://www-files.anthropic.com/production/files/responsible-scaling-policy-1.0.pdf. [Accessed Dec. 15, 2023].
81. "Dario Amodei's Prepared Remarks from the AI Safety Summit on Anthropic's Responsible Scaling Policy," *anthropic.com*, Nov. 1, 2023. [Online]. Available: https://www.anthropic.com/index/uk-ai-safety-summit. [Accessed Dec. 15, 2023].
82. "Infographic: Biosafety Lab Levels," *cdc.gov*, Aug. 30, 2021. [Online]. Available: https://www.cdc.gov/orr/infographics/biosafety.htm. [Accessed Dec. 15, 2023].
83. T. Shevlane, et al., "Model Evaluation for Extreme Risks," arXiv 2023. Available: https://doi.org/10.48550/arXiv.2305.15324.
84. E. J. Hu, et al., "LoRA: Low-Rank Adaptation of Large Language Models," arXiv 2021, Available: https://doi.org/10.48550/arXiv.2106.09685.
85. S. Lermen, C. Rogers-Smith, and J. Ladish, "LoRA Fine-Tuning Efficiently Undoes

Safety Training in Llama 2-Chat 70B," arXiv 2023, Available: https://doi.org/10.48550/arXiv.2310.20624.
86. "Frontier Threats Red Teaming for AI Safety," *anthropic.com*, Jul. 26, 2023. [Online]. Available: https://www.anthropic.com/index/frontier-threats-red-teaming-for-ai-safety. [Accessed Dec. 15, 2023].
87. S. Nevo, et al., "Securing Artificial Intelligence Model Weights: Interim Report," *rand.org*. [Online]. Available: https://www.rand.org/pubs/working_papers/WRA2849-1.html. [Accessed Dec. 15, 2023].
88. J. Leike, "Self-Exfiltration is a Key Dangerous Capability," *aligned.substack.com*, Sep. 13, 2023. [Online]. Available: https://aligned.substack.com/p/self-exfiltration. [Accessed Dec. 15, 2023].
89. E. Seger, et al. "Open-Sourcing Highly Capable Foundation Models: An Evaluation of Risks, Benefits, and Alternative Methods for Pursuing Open-Source Objectives," *Centre for the Governance of AI*, Oct. 10, 2023. [Online]. Available: https://papers.ssrn.com/sol3/papers.cfm?abstract_id=4596436. [Accessed Dec. 15, 2023].
90. "The Power of Open Source AI," *forbes.com*, May 22, 2019. [Online]. Available: https://www.forbes.com/sites/insights-intelai/2019/05/22/the-power-of-open-source-ai/?h=537d8dc46300. [Accessed Dec. 15, 2023].
91. A. Zou, et al., "Universal and Transferable Adversarial Attacks on Aligned Language Models," arXiv 2023, Available: https://doi.org/10.48550/arXiv.2307.15043.